\documentclass[balance,colorlinks]{asmeconf}

\usepackage{amsmath}
\usepackage{cases}
\usepackage{booktabs}
\usepackage{multirow}
\usepackage{listings}
\newcommand{\tabincell}[2]{\begin{tabular}{@{}#1@{}}#2\end{tabular}} 
\usepackage{graphicx} %
\usepackage{makecell}
\usepackage[strings]{underscore}

\hypersetup{pdfauthor={Qineng Wang, Liming Song, Zhendong Guo, Jun Li, Zhenping Feng}, pdftitle={A Novel Multi-Fidelity Surrogate for Turbomachinery Design Optimization}}

\begin{document}
\fancyfoot[RO,RE]{Copyright~\textcopyright~2023 by ASME}

\ConfName{Proceedings of ASME Turbo Expo 2023\linebreak Turbomachinery Technical Conference and Exposition}
\ConfAcronym{GT2023}
\ConfDate{June 26-30, 2023} %
\ConfCity{Boston, Massachusetts, USA} %
\PaperNo{GT2023-104237}

\title{A Novel Multi-fidelity Surrogate for Turbomachinery Design Optimization} %

\SetAuthors{%
	Qineng Wang\affil{1},
    Liming Song\affil{1},
	Zhendong Guo\affil{1}\CorrespondingAuthor{guozhendong@xjtu.edu.cn}, 
	Jun Li\affil{1}, 
    Zhenping Feng\affil{1}  
	}

\SetAffiliation{1}{Institute of Turbomachinery, Xi'an Jiaotong University, Xi'an, China }

\maketitle
\begingroup\renewcommand{\thefootnote}{}
\footnotetext{Published article: Qineng Wang, Liming Song, Zhendong Guo, Jun Li, Zhenping Feng, ``A Novel Multi-Fidelity Surrogate for Turbomachinery Design Optimization,'' Proceedings of ASME Turbo Expo 2023, Volume 13D, V13DT34A023 (2023). DOI: \url{https://doi.org/10.1115/GT2023-104237}. ASME is the original publisher. Copyright \textcopyright\ 2023 ASME.}
\endgroup


\begin{abstract}
The design optimization of turbomachinery is a challenging task as it involves expensive black-box problems. 
The sample-efficient multi-fidelity optimization (MFO) algorithm has been proposed as an efficient solution to this problem. 
By utilizing multi-fidelity surrogates (MFS), the MFO algorithm can use fewer high-fidelity samples aided by low-fidelity samples to establish an accurate surrogate model. 
However, when MFS is used in sequential sampling optimization, it has been observed that the final optimal solution obtained by single-fidelity optimization (SFO) is better than that of MFO, even though MFO performs better at the early stages. 
This can be attributed to the assumption of an even and nested distribution of samples, which is incorrect when using a sequential adding strategy.
To address these issues, we propose a novel algorithm called multi-single-fidelity optimization (MSFO) to overcome the limitations of the conventional MFO procedures. 
In the surrogate establishment of MSFO, we use the density-based spatial clustering of applications with noise (DBSCAN) method to detect local areas where low-fidelity samples are no longer effective. 
A combination of both global MFS and local single-fidelity surrogate model, built using high-fidelity samples alone, is used to establish an ensemble, which improves the anti-interference ability of the algorithm against misleading low-fidelity data.
The effectiveness of the MSFO algorithm is verified first on numerical benchmark functions. Then, the algorithm is used to optimize the aerodynamic profile of a turbine and the film cooling layout design of a turbine endwall. 
Here, high-fidelity sample sources are obtained from fine-mesh CFD simulations, whereas low-fidelity sample sources are obtained from the same simulations run on a coarser mesh. 
The results demonstrate that our MSFO algorithm performs significantly better than the conventional SFO and MFO processes, with a higher level of robustness. 
Therefore, the effectiveness of the MSFO algorithm is well demonstrated.
\end{abstract}
\begin{nomenclature}
    \entry{SFS}{single fidelity surrogate}
    \entry{MFS}{multi-fidelity surrogate}
    \entry{HF}{high-fidelity}
    \entry{MFS}{low-fidelity}
    \entry{MSFO}{multi-single-fidelity fusion optimization}
    \entry{EMFS}{ensemble multi-fidelity surrogate}
    \end{nomenclature}
\section{Introduction}
\par
Computational fluid dynamics (CFD) has been more and more widely used in turbomachinery in the past 20 years~\cite{songResearchMetamodelBasedGlobal2016,ruanVariablefidelityProbabilityImprovement2020}. 
CFD simulations can provide detailed information similar to real physical experiments, which is convenient for global optimization and uncertainty analysis in turbomachinery design problems~\cite{songOptimizationKnowledgeDiscovery2018}. 
High-fidelity CFD simulations require significant computational resources, rendering traditional optimization algorithms unable to complete turbomachinery optimization design. 
The surrogate-based optimization (SBO) algorithm was proposed to address this issue. 
This method employs approximate surrogates to replace expensive simulations in engineering design and achieve efficient optimization, and has been widely applied in various engineering disciplines in recent years.
There are two common methods for SBO: one-shot and adaptive sequential sampling~\cite{liuSurveyAdaptiveSampling2017}. 
The one-shot method involves establishing the surrogate model using a set number of evenly distributed samples, before selecting the optimal design based on the surrogate. 
The adaptive sequential sampling method adds samples one by one and continually improves the surrogate model~\cite{liuSequentialSamplingGeneration2021a}. 
Efficient global optimization (EGO) algorithm~\cite{jonesEfficientGlobalOptimization} is the most typical representative of adaptive sequential sampling SBO algorithm.  
The EGO algorithm's main steps are as follows: 
collect initial samples, 
establish or update the kriging surrogate model, 
and select and add new samples based on the maximum expected improvement (EI) search criteria. 
These steps are repeated until the stop condition is met. 
The adaptive sampling strategy is particularly useful for simulation-based problems. 
By placing more samples in the "region of interest", it is possible to build an accurate global surrogate with fewer points than the one-shot method, thereby reducing the cost of the whole optimization process.
\par
In order to further improve the optimization efficiency, researchers also proposed multi-fidelity optimization (MFO) algorithm~\cite{forresterMultifidelityOptimizationSurrogate2007,makkarMachineLearningFramework2022}, which carries out optimization search based on multi-fidelity surrogate using two or more kinds of data. 
The high-fidelity (HF) data can meet the accuracy requirements of the design tasks, but whose computational cost is expensive. 
The low-fidelity (LF) data is much cheaper, but also inaccurate. 
The accuracy of multi-fidelity surrogate (MFS) with a small amount of HF samples assisted by LF samples is usually significantly better than that of single-fidelity surrogate (SFS) with the same cost~\cite{parkRemarksMultifidelitySurrogates2017,shiMultiFidelityModelingAdaptive2020}. 
Therefore, the MFO algorithm can effectively improve the optimization efficiency of engineering problems with expensive sample evaluation, which has been proved in the design optimization of turbomachinery~\cite{benamaraMultifidelityPODSurrogateassisted2017,mondalMultiFidelityGlobalLocalOptimization2019}.
\par
Compared to single-fidelity optimization, MFO algorithms may converge more quickly but do not necessarily lead to better optimization outcomes~\cite{linSequentialSamplingApproach2022a}. 
Specifically, MFO may lead to inadequate local search capabilities, a phenomenon observed in many previous studies~\cite{guoGenerativeMultiformBayesian2022,wangTransferOptimizationAccelerating2020a}. 
SBO algorithms depend on the accuracy of their surrogate models to determine their search abilities. 
Poor accuracy of the multi-fidelity model in the local optimal region directly leads to inadequate local search abilities in later stages.
\par
Several studies have shown above phenomenon in some cases, adding LF data may not improve the surrogate model's accuracy and may even decrease it below that achieved by using only HF samples in a single-fidelity scheme~\cite{gisellefernandez-godinoIssuesDecidingWhether2019}. 
This phenomenon is referred to as "low-fidelity ineffectiveness" or "low-fidelity reversal"  in this paper, since LF samples fail to improve the surrogate model's accuracy. 
Guo has investigated the relationship between "LF ineffectiveness" and the ratio of HF and LF sample quantities~\cite{guoAnalysisDatasetSelection2018}, 
while Zhou has explored its relationship with the relative accuracy of the LF data~\cite{zhouGeneralizedHierarchicalCoKriging2020}. 
In light of the uneven sample distribution caused by sequential sampling, it is possible for "LF ineffectiveness" to occur only locally when most samples are concentrated in the optimal region. 
Specifically, adding LF samples improves the surrogate model's global accuracy but reduces its accuracy in the local region, ultimately impairing local search abilities as mentioned earlier.
\par
Many researchers also consider the case of "LF ineffectiveness" and propose some novel multi-fidelity models, in order to reduce the influence of LF data when "LF ineffectiveness" occurs, and even directly degrade the model to SFS by adjusting the scale factor $\rho$ in the MFS~\cite{parkLowfidelityScaleFactor2018,shuNovelApproachSelecting2019,buSelectingScaleFactor2022} 
However, these studies are often based on the assumption of uniform distribution of samples and lack the ability to adjust local "LF ineffectiveness" cases. 
Therefore, it is necessary to propose a new MFO algorithm for selecting the single and multi-fidelity surrogate locally in the case of uneven distribution of samples in sequential sampling optimization.

Previous research has addressed the issue of "LF ineffectiveness" through the development of multi-fidelity models. 
These models aim to minimize the effect of LF data when "LF ineffectiveness" occurs and can even recalculate the model to SFS by adjusting the scale factor $\rho$ in the MFS~\cite{parkLowfidelityScaleFactor2018,shuNovelApproachSelecting2019,buSelectingScaleFactor2022}. Nevertheless, these studies are limited by the assumption of a uniform distribution of samples which precludes them from handling local "LF ineffectiveness" cases. 
Therefore, a new MFO algorithm is required to select the single and multi-fidelity surrogate locally and flexibly that can handle unevenly distributed samples during sequential sampling optimization.
\par%
Based on the above views, a novel MFO algorithm, multi-single fidelity fusion optimization algorithm, is proposed, labeled as MSFO. For the uneven sample distribution caused by sequential sampling optimization, MSFO algorithm can adaptive "degenerate" the MFS to the SFS in the local region. To achieve this, the DBSCAN algorithm is used to identify highly uneven "high-density" HF samples; 
Then the MFS and SFS are weighted combination by using the reciprocal of variance method; 
Finally, the obtained ensemble multi-fidelity surrogate (EMFS) will be used to guide subsequent optimization sequence sampling.
\par
The contributions of this paper are mainly reflected in the following aspects: 
(1) A multi-fidelity model for sequential adaptive sampling algorithm is proposed, which fully considers the influence of uneven distribution of samples; 
(2) The local search ability of the proposed MSFO algorithm is effectively improved compared with conventional MFO algorithm; 
(3) The robustness of the MFO algorithm is effectively improved, so that LF data with lower relative accuracy can also be used, and the application scope of the MFO algorithms is expanded.
\par
The following contents of this paper are as follows: 
Section 2 introduces the related work involved in the proposed algorithm; 
Section 3 introduces the steps of the proposed MSFO algorithm in detail, and the effectiveness of the algorithm is preliminarily verified by numerical function examples.
In Section 4, the proposed algorithm is also used to solve two practical engineering design problems;
Finally, in Section 5, a brief conclusion is made about this work.
\section{Research background}
The fundamental concept behind MSFO is to identify the likelihood of "LF ineffectiveness" in a region based on the sample density and to assign a weight coefficient to each coordinate. The EMFS in the MSFO algorithm should be created by merging SFS and MFS using those weight coefficients. 
In the above procedure, the identification of high-density regions is accomplished using the DBSCAN method. 
Kriging and co-kriging surrogates, respectively, are the SFS and MFS techniques applied in this study.
\subsection{DBSCAN algorithm}
\par
The DBSCAN algorithm is a density-based clustering algorithm in cluster analysis~\cite{schubertDBSCANRevisitedRevisited2017}. 
Its basic idea is that every point in the cluster contains at least a given number of points $\text{minPts}$ within a given radius $\varepsilon$. The algorithm classifies regions with sufficient density into a class.
For efficiency reasons, DBSCAN does not perform density estimation in-between points. 
Instead, all neighbors within the radius $\epsilon$ of a core point are considered to be part of the same cluster as the core point. 
If any of these neighbors is again a core point, their neighborhoods are also transitively included. 
All points within the same set are density connected. 
Points which are not density reachable from any core point are considered not belong to any cluster.
\par
\begin{figure}[hbpt]
    \centering\includegraphics[width=0.75\linewidth]{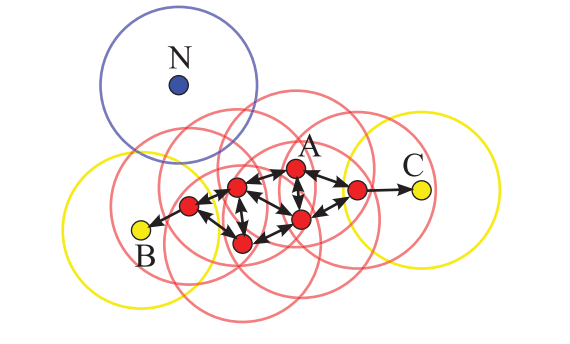}
    \caption{Illustration of the DBSCAN clustering~\cite{schubertDBSCANRevisitedRevisited2017}. \label{fig:1}}
    \end{figure}

Figure \ref{fig:1} illustrates the concepts of DBSCAN. The $\text{minPts}$ parameter is 4, and the radius $\epsilon$ is indicated by the circles. 
\subsection{Kriging surrogate}
\par
Kriging is a popular surrogate technique~\cite{forresterRecentAdvancesSurrogatebased2009}.
The kriging prediction $Y_{krg}$ at unknown site $x$ is built as a trend function $f(x)$ plus a normal random process $Z(x)$ as:
\begin{equation}\label{eq:2.1}
{Y_{krg}}({\bf{x}}) = f({\bf{x}}) + Z({\bf{x}})
\end{equation}
where, $f(x)$ is usually a constant; $Z(x)$ describes the local features of $Y$ around the $n$ sample points.
The function prediction at an unknown point $\bf{x}$ can be expressed as:
\begin{equation}\label{eq:2.2}
{{\hat y}_{krg}}({\bf{x}}) = \hat \mu  + {{\bf{r}}^T}{{\bf{R}}^{ - 1}}({\bf{y}} - {\bf{1}}\hat \mu )\\
\end{equation}
where $\mu$ is the regression constant as 
$\mu=\left(
\mathbf{1}^{T} \mathbf{R}^{-1} \mathbf{1}
\right)^{-1} 
\mathbf{1}^{T} \mathbf{R}^{-1} \mathbf{y}_{S}$;
$\mathbf{R}$ is the correlation matrix, and $\mathbf{r}$ is the correlation vector.
More details about the establish of kriging surrogate can be found in Ref.~\cite{jonesTaxonomyGlobalOptimization}.
\subsection{Co-kriging surrogate}
\par 
Co-kriging is a surrogate technique that is extended base on kriging. 
The difference is that it can leverage samples from multi-fidelity sources to form one surrogate~\cite{kennedyPredictiwnghetnheFaOsuttApuptprforoxmimaatCioonmsParleexACvoamilapbulteerCode}. 
\par 
 A crucial assumption expressed in Eq.(\ref{eq:2.3}) is used for establish a connection between HF and LF data.  
\begin{equation}\label{eq:2.3}
y_{H}({\bf{x}})=\rho y_{L}({\bf{x}})+Z_{d}({\bf{x}})
\end{equation}
where $\rho$ is the scale factor, $Z_{d}({\bf{x}})$ is called discrepancy function (DF).%
And the scale factor $\rho$ also plays a role in adjusting the weight of LF data. 
\par Finally, the predictor of co-kriging is in Eq.(\ref{eq:2.4}), where $C$ and $c$ are the correlation matrix and the correlation vector in co-kriging surrogate.
\begin{equation}\label{eq:2.4}
\hat{y}_{h}({\bf{x}}^{*})=\beta_{0}+c^{T}({\bf{x}}^{*})C^{-1}(y_{s}-\beta_{0}F)\\
\end{equation}
More details about the establish of co-kriging surrogate can be found in Ref.~\cite{forresterMultifidelityOptimizationSurrogate2007}.
\section{Proposed Method}
\par
\begin{figure}[ht]
\centering\includegraphics[width=1\linewidth]{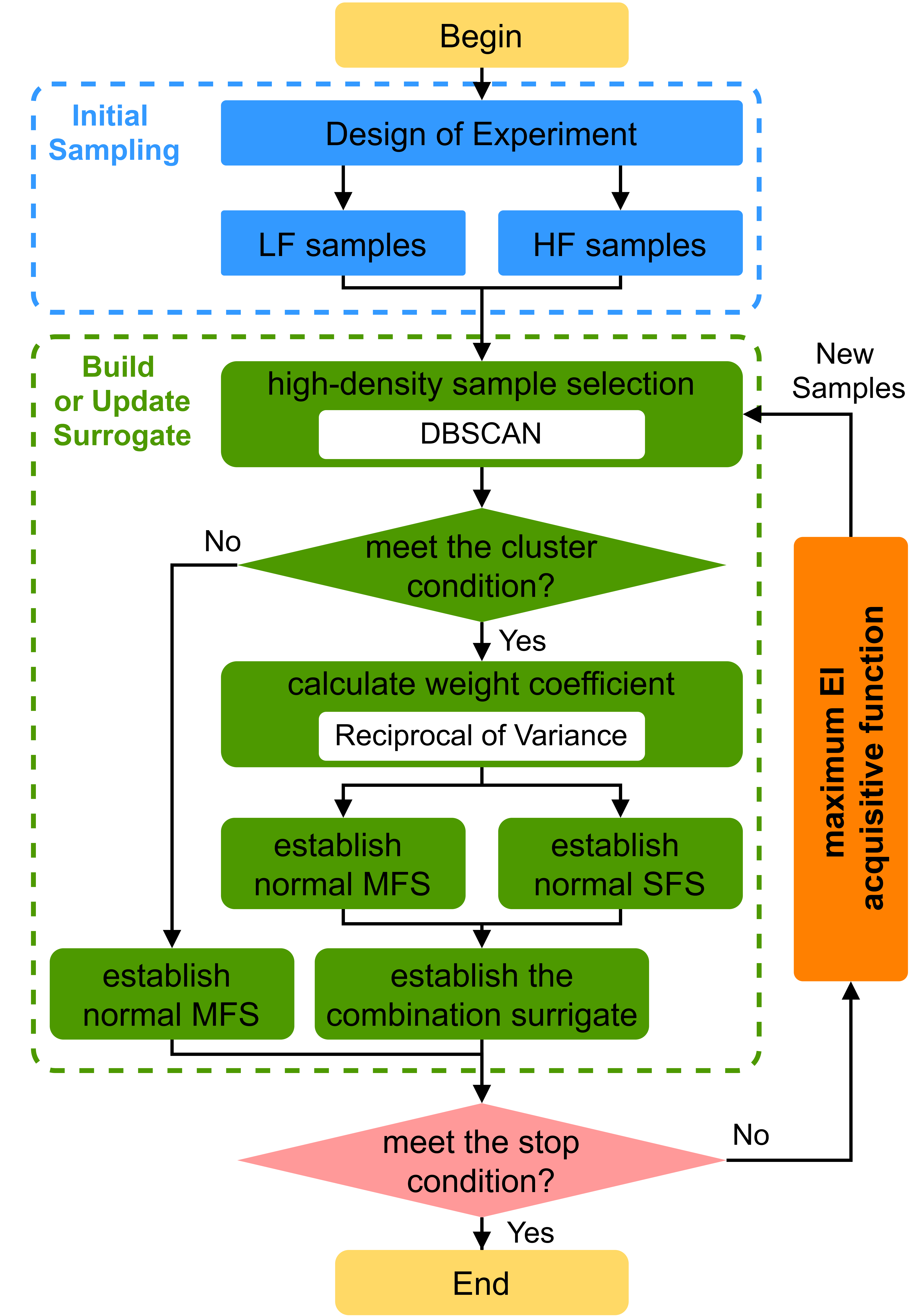}
\caption{The flowchart of the proposed MSFO algorithm \label{fig:2}}
\end{figure}

In order to improve the local searching ability of MFO algorithm, this paper proposes the MSFO algorithm, whose basic framework process is shown in Fig.\ref{fig:2}. 
As can be seen, the biggest difference between MSFO and conventional MFO algorithm is reflected in the step of establishing the surrogate model. 
Beside, in terms of initial sampling, sequencial sampling, and termination conditions, MSFO algorithm is similar to conventional MFO algorithm, It also uses the most widely used maximum EI criteria~\cite{liuSurveyAdaptiveSampling2017} as the acquisitive function as below:
\begin{equation}\label{eq:3.1}
\begin{split}
EI(\bm{x}) =& \begin{cases}
(f_{\text{min}}-\hat{y}(\bm{x}))\Phi(u(\bm{x}))+s(\bm{x})\phi(u(\bm{x}))  \quad (s  \,\textgreater\,\,0)\\
0  \qquad (s\leq 0)
\end{cases} \\
u(\bm{x})=&{(f_{\text{min}}-\hat{y}(\bm{x}))}/{s(\bm{x})}\\
\end{split}
\end{equation}
where, $\Phi \left(  \cdot  \right)$ and $\phi \left(  \cdot  \right)$ denote the standard normal distribution and density functions.
\par
In this paper, an innovative ensemble multi-fidelity surrogate, EMFS, is proposed. 
In the following section, we will describe the fundamental concept and specific implementation steps of this surrogate in detail.
\subsection{Ensemble multi-fidelity surrogate}
\par
Throughout the optimization process, the sequential sampling method intensifies the uneven sample distribution, causing one or more areas of "LF ineffectiveness" to appear in the design space. 
The fundamental approach of the MSFO method is to detect these regions and decrease the weight of the MFS in those locations. 
In perfect circumstances, sequential sampling would cause the weight of SFS in the "LF ineffectiveness" regions to gradually rise, leading to the gradual expansion of the areas recognized as "LF ineffectiveness". 
As concentrated sample distribution is typically limited to the local optimal region, we will refer to this region as the "LF ineffectiveness" region throughout the paper. 
To achieve these objectives, this paper implements the ensemble surrogate method, as depicted in Eq.(\ref{eq:3.2}).
\begin{equation}\label{eq:3.2}
\begin{array}{c}
    {{\hat y}^{\text{EMFS}}}({\bf{x}}) = {p_G}({\bf{x}}) \cdot \hat y_G^{ck}({\bf{x}}) + {p_L}({\bf{x}}) \cdot \hat y_L^{krg}({\bf{x}})\\
    {{\hat s}^{\text{EMFS}}}({\bf{x}}) = {p_G}({\bf{x}}) \cdot \hat s_G^{ck}({\bf{x}}) + {p_L}({\bf{x}}) \cdot \hat s_L^{krg}({\bf{x}})
    \end{array}
\end{equation}
where, the subscript $G$ and $L$ are represent the global MFS and the local SFS respectively.
\par
The EMFS requires the establishment of SFS and MFS at the same time, and gives different weights to them in different regions. 
Combined with the maximum EI criteria, the number of HF samples in "LF ineffectiveness" region keeps increasing, which increases the weight of the SFS in this region, effectively avoiding the accuracy reduction caused by the "LF ineffectiveness".
\par
The establishment of the EMFS can be divided into three steps: 
(1) Identification of high-density HF samples; 
(2) Determine the weight coefficient in the design space;
(3) Establishment of the ensemble surrogate. 
These steps are described in detail below.
\subsubsection{Identification of high-density HF samples}
\par
In constructing the EMFS, each HF sample is classified into two types: "local high-density sample" and "global low-density sample". 
This classification is used to determine the weight of the model for each coordinate position. 
Since there are sufficient HF samples located in closed proximity to the "locally high-density samples", LF samples cannot provide additional valuable information, rendering them located within the "LF ineffectiveness" region. 
Conversely, the "low-density sample" is surrounded by an absence of samples, which necessitates the incorporation of additional LF samples to enhance the accuracy. 
The classification of HF samples is completed using the DBSCAN method based on the concept of "density".
\par
As mentioned above, the execution of DBSCAN algorithm requires the user to set the clustering radius $\varepsilon$ and the minimum number of clusters $\text{minPts}$. The parameters shown in Eq.(\ref{eq:3.3}) are adopted in this paper.
\begin{equation}\label{eq:3.3}
\begin{array}{l}
\varepsilon  = {d^{ - 1/2}}{n_H}^{ - 1/d}\\
\text{minPts} = d
\end{array}
\end{equation}
where $d$ represents the problem dimension and $n_{H}$ represents the number of HF samples.
\subsubsection{Determine the weight coefficient in the design space}
\par
The weight parameter determines the ratio of SFS and MFS weight values for each coordinate position within the design space which, in turn, affect the accuracy of the final model. 
The reciprocal of variance weighting method is commonly used to determine the weighting coefficient between different models~\cite{zerpaOptimizationMethodologyAlkaline2005}. 
The smaller the estimated variance of the model at the coordinate $x$, the higher its accuracy. 
Therefore, it is essential to allocate a larger weight value to the model in that area. 
Based on these concepts, the weight coefficient calculation method in the ensemble multi=fidelity surrogate (EMFS) can be expressed as follows:
\begin{footnotesize}
\begin{equation}\label{eq:3.4}
\begin{array}{c}
    {\bf{R}} = \left[ {\begin{array}{*{20}{c}}
    {R\left( {{{\bf{x}}^{(1)}},{{\bf{x}}^{(1)}}} \right)}& \cdots &{R\left( {{{\bf{x}}^{(1)}},{{\bf{x}}^{(n)}}} \right)}\\
     \vdots &{}& \vdots \\
    {R\left( {{{\bf{x}}^{(n)}},{{\bf{x}}^{(1)}}} \right)}& \cdots &{R\left( {{{\bf{x}}^{(n)}},{{\bf{x}}^{(n)}}} \right)}
    \end{array}} \right] = \left[ {\begin{array}{*{20}{c}}
    {{{\bf{R}}_{GG}}}&{{{\bf{R}}_{LG}}}\\
    {{{\bf{R}}_{GL}}}&{{{\bf{R}}_{LL}}}
    \end{array}} \right]\\ %
    {\bf{r}}({\bf{x}}) = \left[ {\begin{array}{*{20}{c}}
    {R\left( {{{\bf{x}}^{(1)}},{\bf{x}}} \right)}\\
     \vdots \\
    {R\left( {{{\bf{x}}^{(n)}},{\bf{x}}} \right)}
    \end{array}} \right] = \left[ {\begin{array}{*{20}{c}}
    {{{\bf{r}}_G}({\bf{x}})}\\
    {{{\bf{r}}_L}({\bf{x}})}
    \end{array}} \right] %
    \end{array}
\end{equation}
\end{footnotesize}
\par
As shown in Eq.(\ref{eq:3.4}), all HF samples are used to establish the kriging surrogate, and its relational matrix $\bf{R}$ and relational vector $\bf{r}$ can be obtained. ${\bf{R}}_{GG}$ and ${\bf{R}}_{LL}$ represent the internal relations of the two types of samples respectively. 
The variance of each type of sample in arbitrary coordinates can be independently estimated through the block matrix, as shown in Eq.(\ref{eq:3.5}).
\begin{small}
\begin{equation}\label{eq:3.5}
    \begin{array}{c}
        {p_L}({\bf{x}}) = {\rm{sigmoid}}\left( 
            {\hat v({{\bf{R}}_{GG}},{{\bf{r}}_G},{\bf{x}})}/
            ({{\hat v({{\bf{R}}_{GG}},{{\bf{r}}_G},{\bf{x}}) + \hat v({{\bf{R}}_{GG}},{{\bf{r}}_G},{\bf{x}})}})
            \right)\\

     \hat{\mathbf{v}}(\mathbf{x}, \mathbf{R}, \mathbf{r})=1-\mathbf{r}^{\mathrm{T}} \mathbf{R}^{-1} \mathbf{r}+\left(1-\mathbf{F}^{\mathrm{T}} \mathbf{R}^{-1} \mathbf{r}\right)^2 / \mathbf{F}^{\mathrm{T}} \mathbf{R}^{-1} \mathbf{F}\\
        {\rm{sigmoid}}(z) = {1}/{{1 + {e^{ - (mz - b)}}}},{\rm{     }}{p_G}({\bf{x}}) = 1 - {p_L}({\bf{x}})
        \end{array}
    \end{equation}
\end{small}
\par
Finally, the weight coefficient obtained after through a sigmoid function to make the integration weight coefficient change more gently and evenly in the space.
Here, the values of $m$ and $b$ are set as $10$ and $5$.
\subsubsection{Establishment of the ensemble surrogate} 
\par
After the weight coefficient ${p_G}({\bf{x}})$ and ${p_L}({\bf{x}})$ are determined, the SFS and the MFS need to be constructed respectively and weighted to obtain the ensemble surrogate. 
In this paper, the SFS is a kriging surrogate constructed using "high-density samples", while the MFS is a co-kriging surrogate. 
Finally, the ensemble surrogate is created as shown in Eq.(\ref{eq:3.2}). It can be seen from this step that the construction of EMFS does not need to use the special properties of the two weighted models, so this method can be easily extended to other surrogates.
\subsection{A 2-D illustrative example of the EMFS}
\par
\begin{figure}
    \begin{subfigure}[t]{0.25\textwidth} %
    \centering{
    \includegraphics[width=0.7\linewidth]{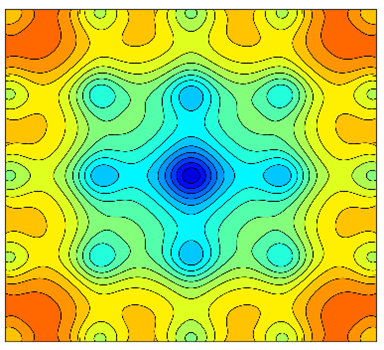}
    }%
    \subcaption{Real function}
    \end{subfigure}%
    \begin{subfigure}[t]{0.25\textwidth}
    \centering{%
    \includegraphics[width=0.8\linewidth]{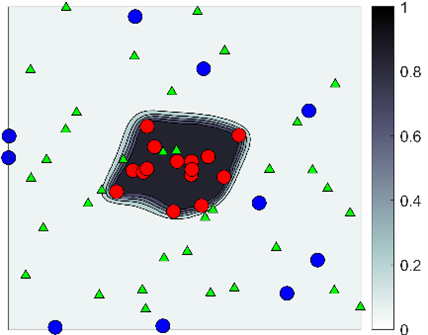}
    }%
    \subcaption{The weight coefficient of SFS}
    \end{subfigure}
    
    \begin{subfigure}[t]{0.5\textwidth} %
    \centering{
    \includegraphics[width=1\linewidth]{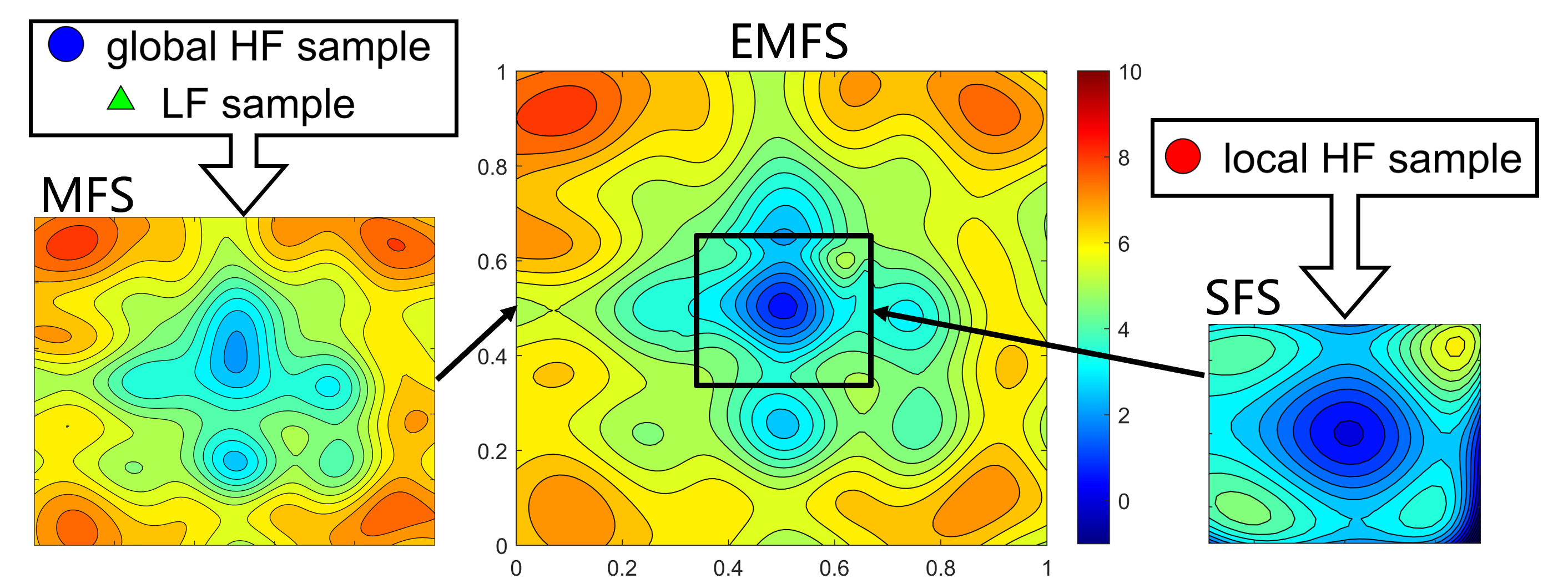}
    }%
    \subcaption{The ensemble multi-fidelity surrogate}
    \end{subfigure}%
    \caption{The illustration of construct process of the EMFS. }\label{fig:3}
    \end{figure}

In order to more specifically display the construction process of the EMFS, 
the diagram of how to construct a EMFS for two-dimensional Ackley function is shown in Fig.\ref{fig:3}. 
The $f_H({\bf{x}})$ and $f_L({\bf{x}})$ functions are expressed as Eq.(\ref{ackley}), and the range of $x_i$ is [-2,2].
The real function of $f_H({\bf{x}})$ is shown in Fig.\ref{fig:3}(a)
\begin{equation}\label{ackley}
\begin{array}{l}
{f_H({\mathbf{x}})} =   - 20\exp \left[ {\sqrt {\frac{1}{n}\sum\limits_{i = 1}^2 {{x_i}^2} } } \right]
- \exp \left[ {\frac{1}{n}\sum\limits_{i = 1}^2 {\cos ({x_i}/\pi)} } \right] + 39.09\\

f_L({\bf{x}}) = f_H({\bf{x}}) + 6.8(0.585 - 0.00127{x_1} + 0.00113{x_2} )\\

\end{array}
\end{equation}
\par
Figure \ref{fig:3}(b) shows the distribution of the samples used, including 40 LF samples $\{X_{L},Y_{L}\}$ and 25 HF samples $\{X_{H},Y_{H}\}$. 
The distribution of HF samples comes from an EGO process, which can effectively simulate the uneven distribution caused by sequential sampling. 
First, as shown in Fig.\ref{fig:3}(b), the DBSCAN algorithm adaptive identifies high density HF samples $X_{H,local}$, which are represented in red circles. 
Then, the weight coefficient ${p_G}({\bf{x}})$ and ${p_L}({\bf{x}})$ are calculated by using Eq.(\ref{eq:3.4}) and Eq.(\ref{eq:3.5}), and also drawn in Fig.\ref{fig:3}(b).
It can be found that, the distribution of weight coefficient ${p_L}({\bf{x}})$ is highly correlated with the distribution of samples $X_{H,local}$. 
And in the "LF ineffectiveness" region, the value of weight coefficient ${p_L}({\bf{x}})$ is close to 1, which means this region is degenerate to SFS locally.
Finally, as shown in Fig.\ref{fig:3}(c), after establishing two kinds of surrogate respectively, the EMFS is obtained by weight addition. 
Globally, the model is similar to the MFS, but in local "LF ineffectiveness" region, the SFS is used to improve the details.
\par
The comparison above illustrates that effectively improving the accuracy of the local region without reducing the accuracy of the global scope can be achieved by degenerating the local region with sufficient HF samples to the SFS. 
This is an ability that cannot be attained by other global-adjusted MFO methods.
\subsection{A numerical test case of the MFSO algorithm}
\par
To compare the proposed MSFO algorithm with single fidelity EGO and conventional co-Kriging optimization (CKO), a 5-dimensional benchmark function was utilized.
\par
In the test cases, the initial sampling is obtained with Latin hypercube sampling (LHS), the number of initial HF samples is $3d$, the number of initial LF samples is $10d$, where $d$ is the function's dimension. 
\par
The 5-D Ackley function is set as the HF function $f_H(\bf{x})$ in this numerical test~\cite{huangSequentialKrigingOptimization2006}.
Then, in order to further prove the anti-interference ability of the MSFO, the LF functions are defined as $f_L({\bf{x}}) = f_H({\bf{x}}) + 6.8 \cdot P\cdot MA5({\bf{x}})$.
The smaller the $P$ value, the closer the LF data is to the HF data.
The Pearson's coefficient~\citep{bertholdClusteringTimeSeries2016} is used to intuitively quantify the degree of similarity between high and low precision data, and its calculation method is as follows:
\begin{equation}\label{eq:4.1}
    corr(A,B)=\frac{\sum_{i=1}^n\left(A_{i}-\bar{A}\right)\left(B_{i}-\bar{B}\right)}{\left\{\sum_{i=1}^n\left(A_{i}-\bar{A}\right)^2 \sum_{j=1}^n\left(B_{j}-\bar{B}\right)^2\right\}^{1 / 2}},
\end{equation}
In the optimization cases, functions with $P$ values of 1, 0.6 and 0.1 are selected as the LF functions, and their Pearson coefficients are 0.33, 0.57 and 0.87, respectively.
Here, every case repeat the calculation for 20 times. 
\par
\begin{figure*}
\begin{subfigure}[t]{0.33\textwidth} %
\centering{
\includegraphics[width=0.9\linewidth]{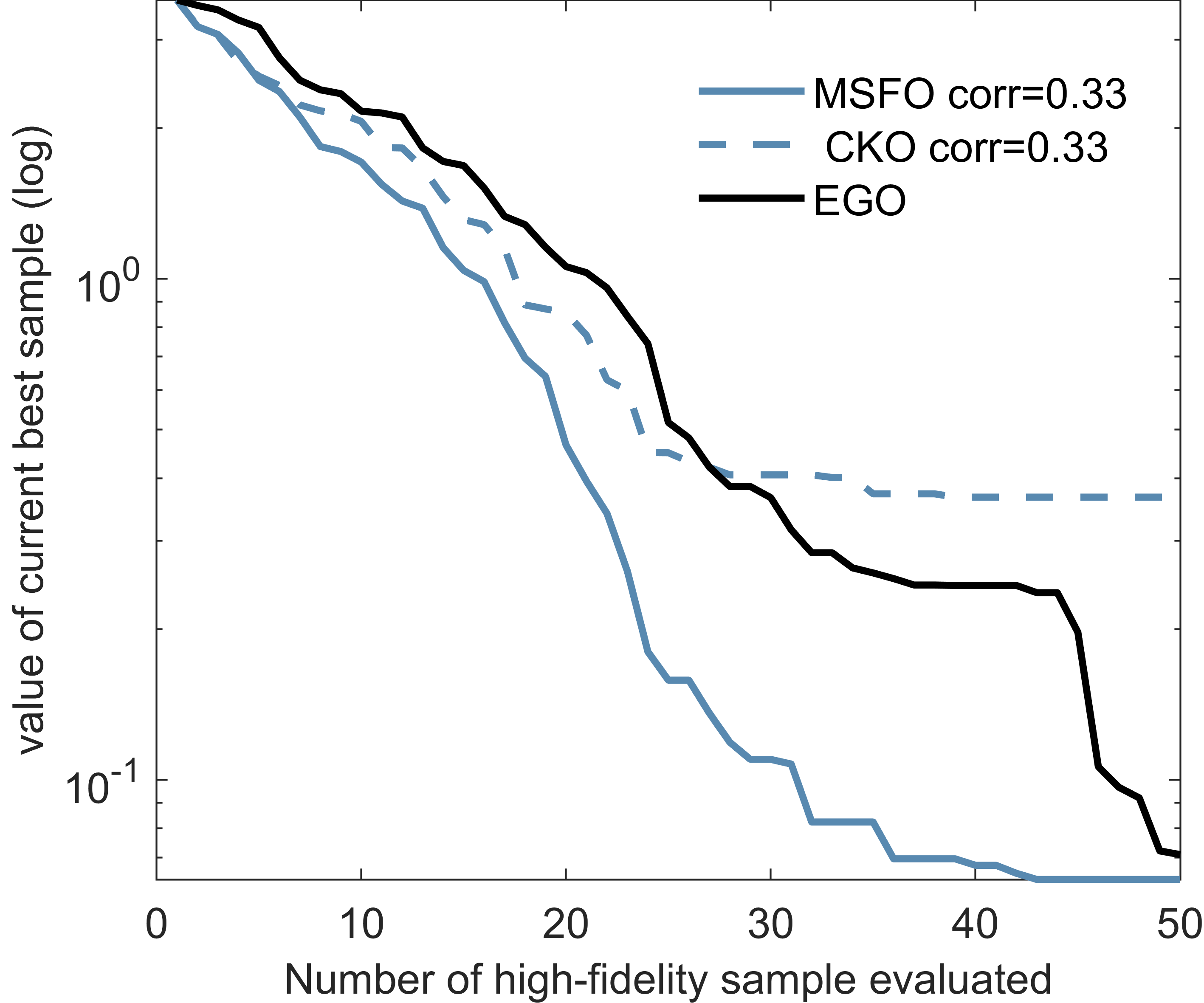}
}%
\subcaption{}
\end{subfigure}%
\begin{subfigure}[t]{0.33\textwidth}
\centering{%
\includegraphics[width=0.9\linewidth]{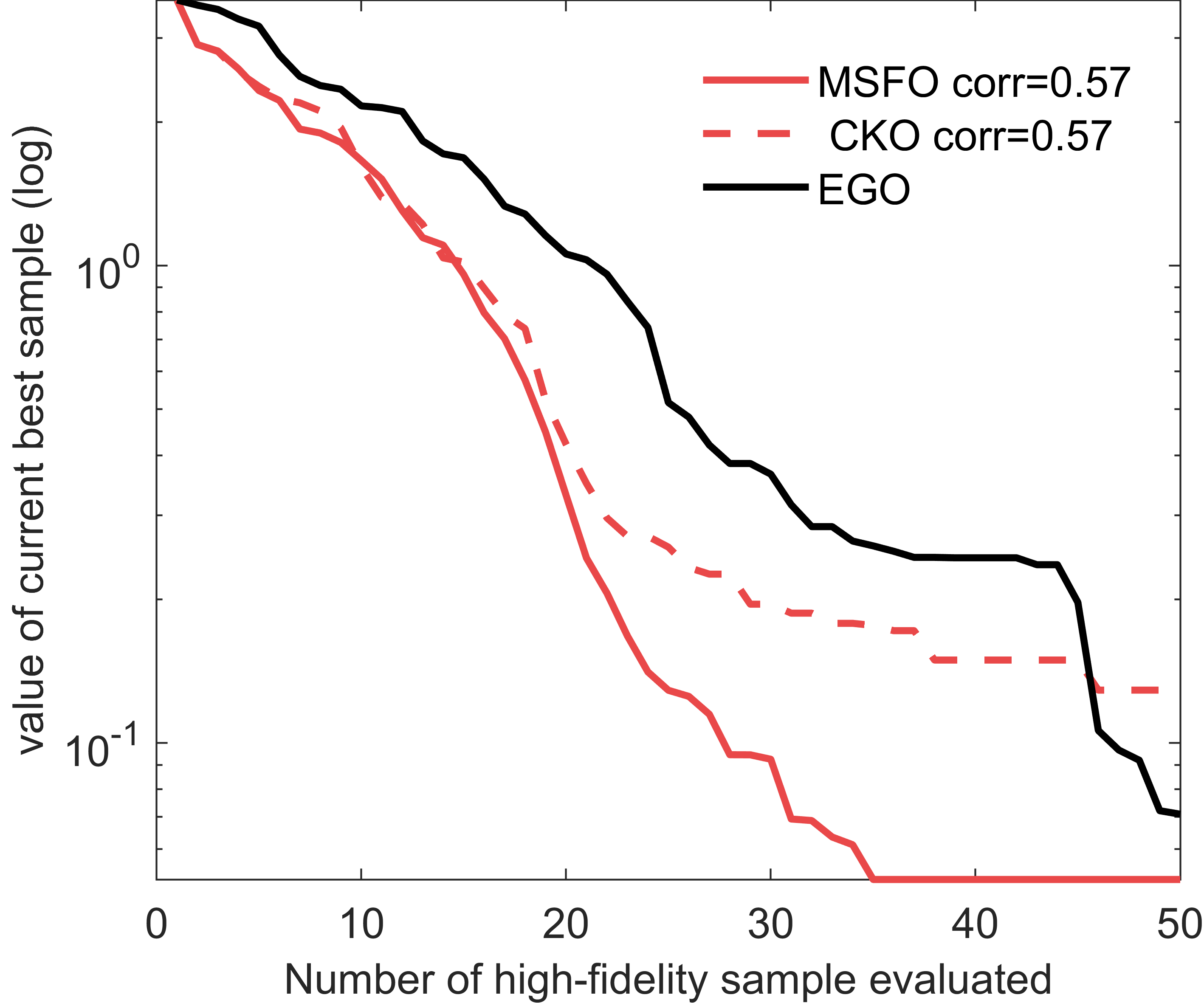}
}%
\subcaption{}
\end{subfigure}
\begin{subfigure}[t]{0.33\textwidth}
\centering{%
\includegraphics[width=0.9\linewidth]{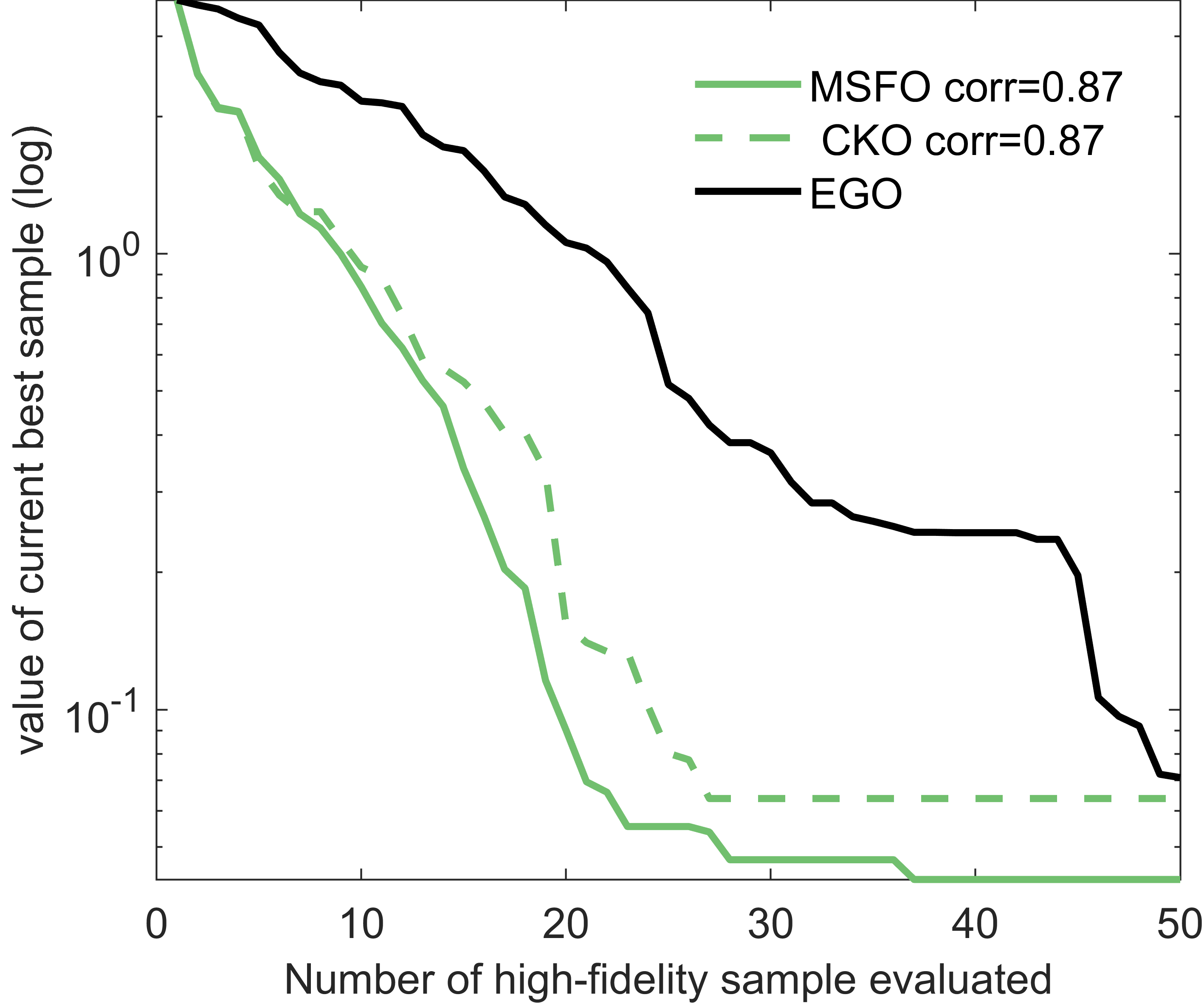}
}%
\subcaption{}
\end{subfigure}

\caption{The average convergence curves of 5-D Ackley function}\label{fig:4}
\end{figure*}

 
The optimization results with different Low-Fidelity (LF) data are presented in Figure \ref{fig:4}, which indicates that the MSFO algorithm performs better in all three examples. 
As shown in Figure \ref{fig:4}(a), when the similarity between High-Fidelity (HF) and LF is 0.33, the LF data is not effective in guiding the optimization search, resulting in worse performance from the CKO algorithm compared to the EGO algorithm. 
In contrast, the MSFO algorithm consistently maintains a faster convergence rate across all examples, demonstrating its effectiveness.

\section{Engineering test cases}
\par
In this section, a 7-D turbine profile aerodynamic optimization problem is used to demonstrate the practicability of the MSFO method in practical engineering tasks. 
Finally, the MSFO has also completed a typical expensive black box engineering problem, a 5-D turbine endwall cooling layout design problem, proving that it also has advantages over state-of-the-art algorithms in tasks with very limited HF samples.
\par
Similar to the above numerical test case in Section 3.3, in engineering optimizations, the initial sampling is obtained with Latin hypercube sampling (LHS), the number of HF samples is $3d$, the number of LF samples is $10d$. 
And the MSFO algorithm is also compared with the EGO and CKO algorithms in the two engineering design task.
\subsection{Turbine blade design problem}
\par
\begin{figure}[htbp]
\centering\includegraphics[width=0.7\linewidth]{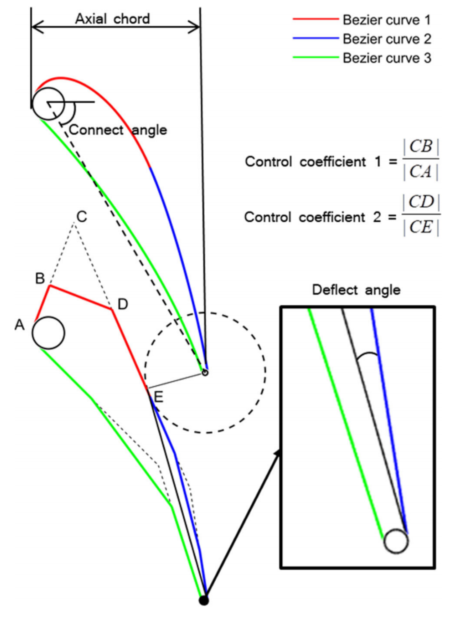}
\caption{The definition of parameter in the optimization of GE-E3 blade \label{fig:5}}
\end{figure}

To validate the applicability of the proposed approach to real-world design problems, a turbine blade parametric design problem is chosen to compare the EMFS and the current approach. 
The first stage turbine guide vane of Energy Efficient Engine is selected as the baseline blade profile to be redesigned. 
The blade profile is defined with 11 geometric parameters and 7 of them are chosen as design variables. 
\begin{table}[htbp]
    \caption{Design variables in the GE-E3 aerodynamic design case}
    \begin{center}
    \renewcommand\arraystretch{1.2}
    \begin{tabular}{cccc}
    \hline
        \tabincell{c}{No.} &  \tabincell{c}{Geometric\\ definition} & \tabincell{c}{Reference\\value }& \tabincell{c}{Variation\\ range} \\
    \hline
            1 & axial chord &      33.9[mm]  &     (-2.0, 2.0) \\
            2 & center connect angle &       59.8[$^\circ$]  &     (-4.0, 2.0) \\
            3 & inlet wedge angle &         69.0[$^\circ$] &    (-15.0, 3.0) \\
            4 & outlet deflect angle &        4.5[$^\circ$]  & (-1.5, 4.5) \\
            5 & correlation coefficient &       0.35[-]  & (-0.05, 0.10) \\
            6 & control coefficient 1 &        0.40[-] & (-0.15, 0.15) \\
            7 & control coefficient 2 &        0.50[-]  & (-0.15, 0.15) \\
    \hline
    \end{tabular}  
    \end{center}
    \end{table}
To illustrate the parameterization method clearly, Fig.\ref{fig:5} presents the definition of several key parameters. 
\par
To measure the aerodynamic performance of the designed blade profile, The energy loss coefficient is defined as the objective of the optimization, which is expressed as below: 
\begin{small}
\begin{equation}\label{eq:4.2}
\begin{split}
&{\bf{x}}_{\text{optimal}} = \text{arg min}  \{\varepsilon({\bf{x}})\}\\
\text{where, }\varepsilon &  = 1 - \left[ {1 - {{\left( {\frac{{{P_{{\rm{outlet }}}}}}{{P_{{\rm{outlet }}}^*}}} \right)}^{\frac{{\gamma  - 1}}{\gamma }}}} \right]/\left[ {1 - {{\left( {\frac{{{P_{{\rm{outlet }}}}}}{{P_{{\rm{inlet }}}^*}}} \right)}^{\frac{{\gamma  - 1}}{\gamma }}}} \right]
\end{split}
\end{equation}
\end{small}
where $P_{\rm{outlet}}$ and  $P_{{\rm{outlet }}}^*$ denote the static and total pressure at the cascade exit respectively. 
In house software is employed to compute the energy loss coefficient.
For more details on design variables and boundary conditions, please see the Ref.~\cite{guoParallelMultifidelityExpected2021}.
\par
\begin{figure}[htbp]
\centering\includegraphics[width=0.7\linewidth]{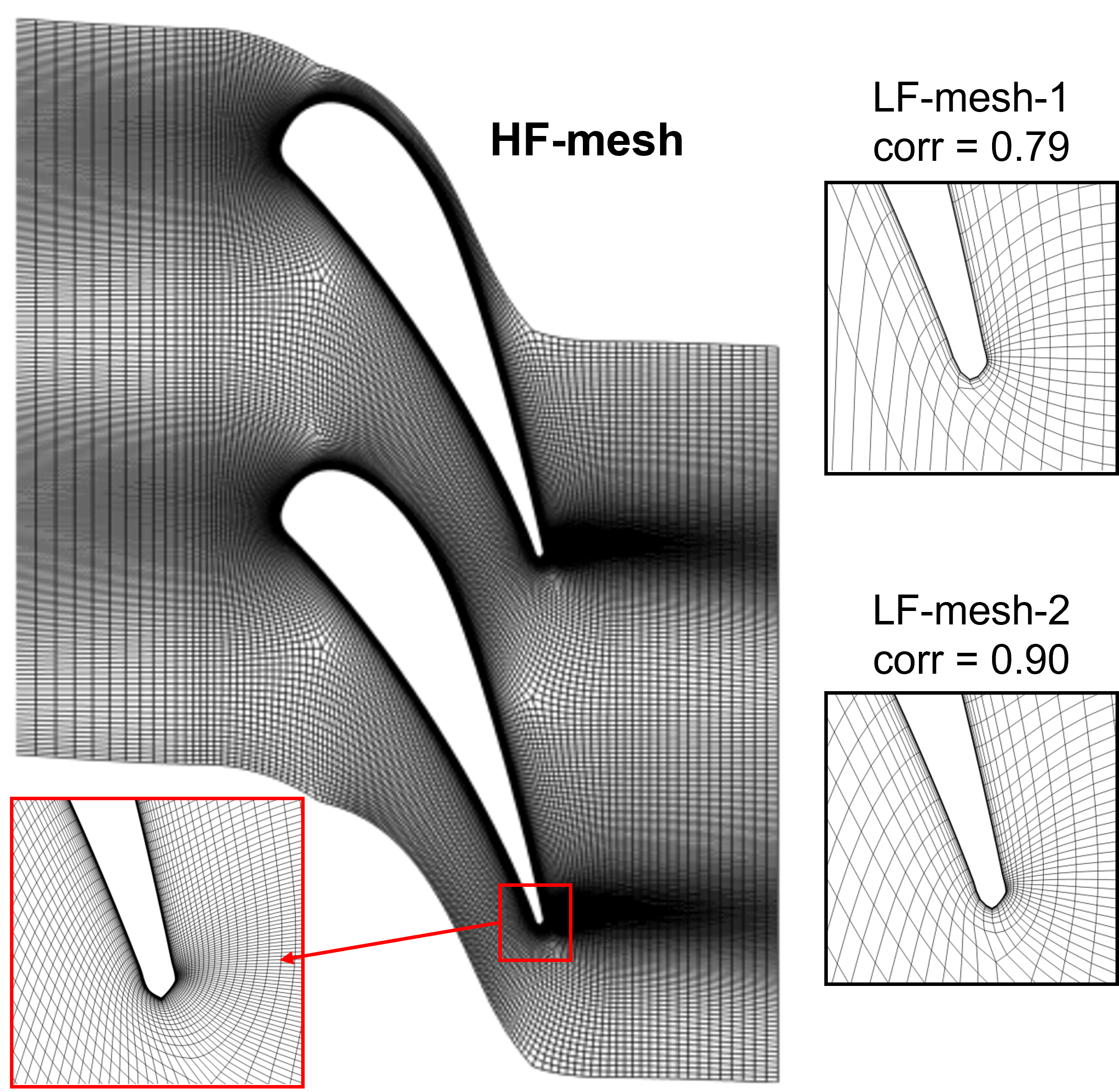}
\caption{Meshes for high- and low-fidelity computations of the GE-E3 aerodynamic design.\label{fig:6}}
\end{figure}

As widely adopted in aeronautical multi-fidelity design, the HF and LF data are obtained from CFD of fine and coarse meshes. The HF-mesh with 23377 nodes are shown in the Fig.\ref{fig:6}, and the details of two LF-meshes with 2890 and 1155 nodes are both shown here, respectively. 
Every case repeat the calculation for 10 times.
And the Pearson coefficient in Fig.\ref{fig:6} are calculated by all 210 initial samples in 10 times repetitions.
\par
\begin{figure}[htbp]
\centering\includegraphics[width=0.7\linewidth]{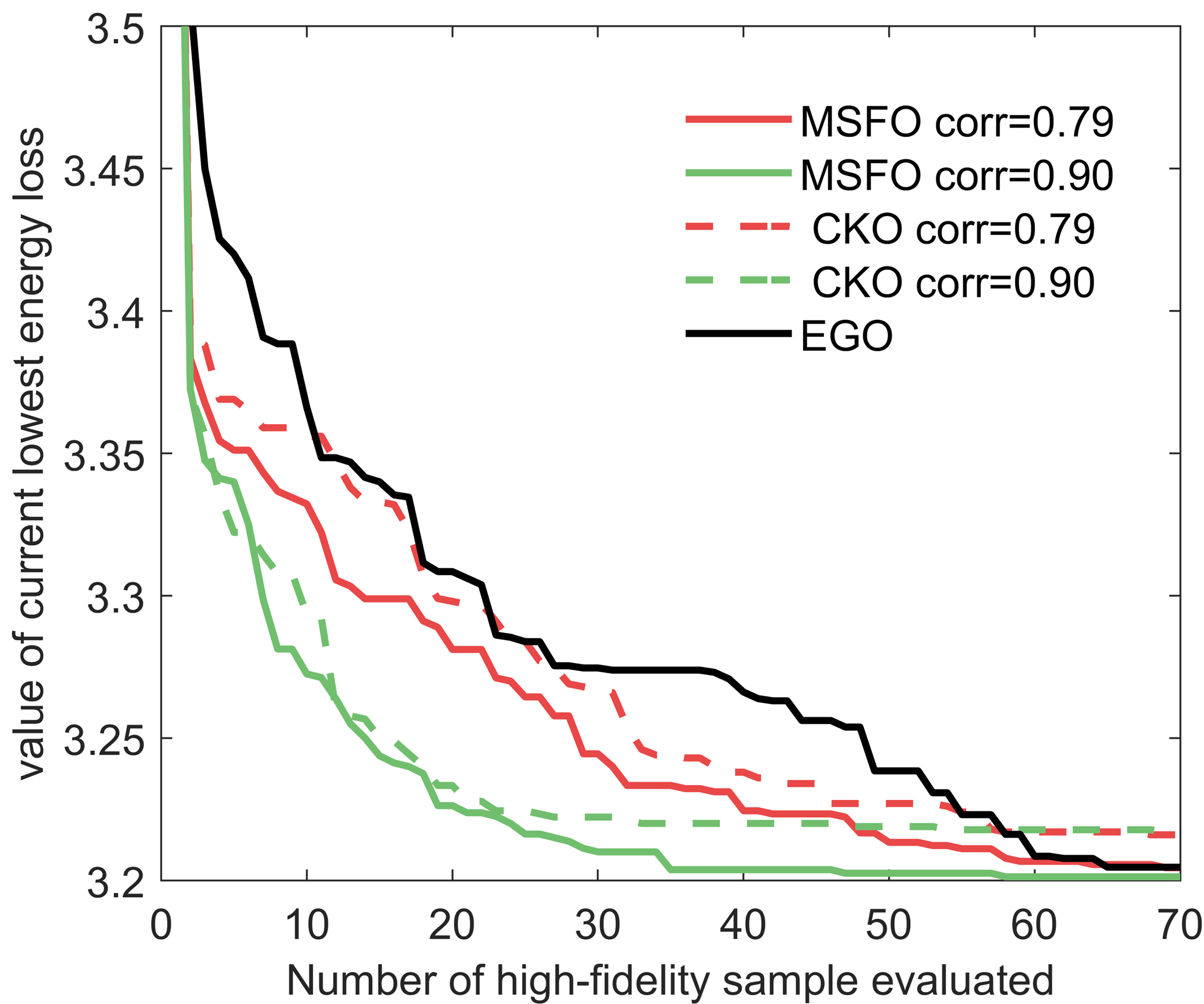}
\caption{The average convergence curves of GE-E3 optimization task\label{fig:7}}
\end{figure}

The optimization result obtained by set different LF sources are shown in Fig.\ref{fig:7}.
In the optimization results of turbine blade design, many phenomena are very similar to those of ackley function optimization in section 3.3. 
The EMFS method shows strong local search ability, and the optimization results are obviously superior to CKO in all cases with different LF data.
\par
\begin{figure}[htbp]
\centering\includegraphics[width=0.7\linewidth]{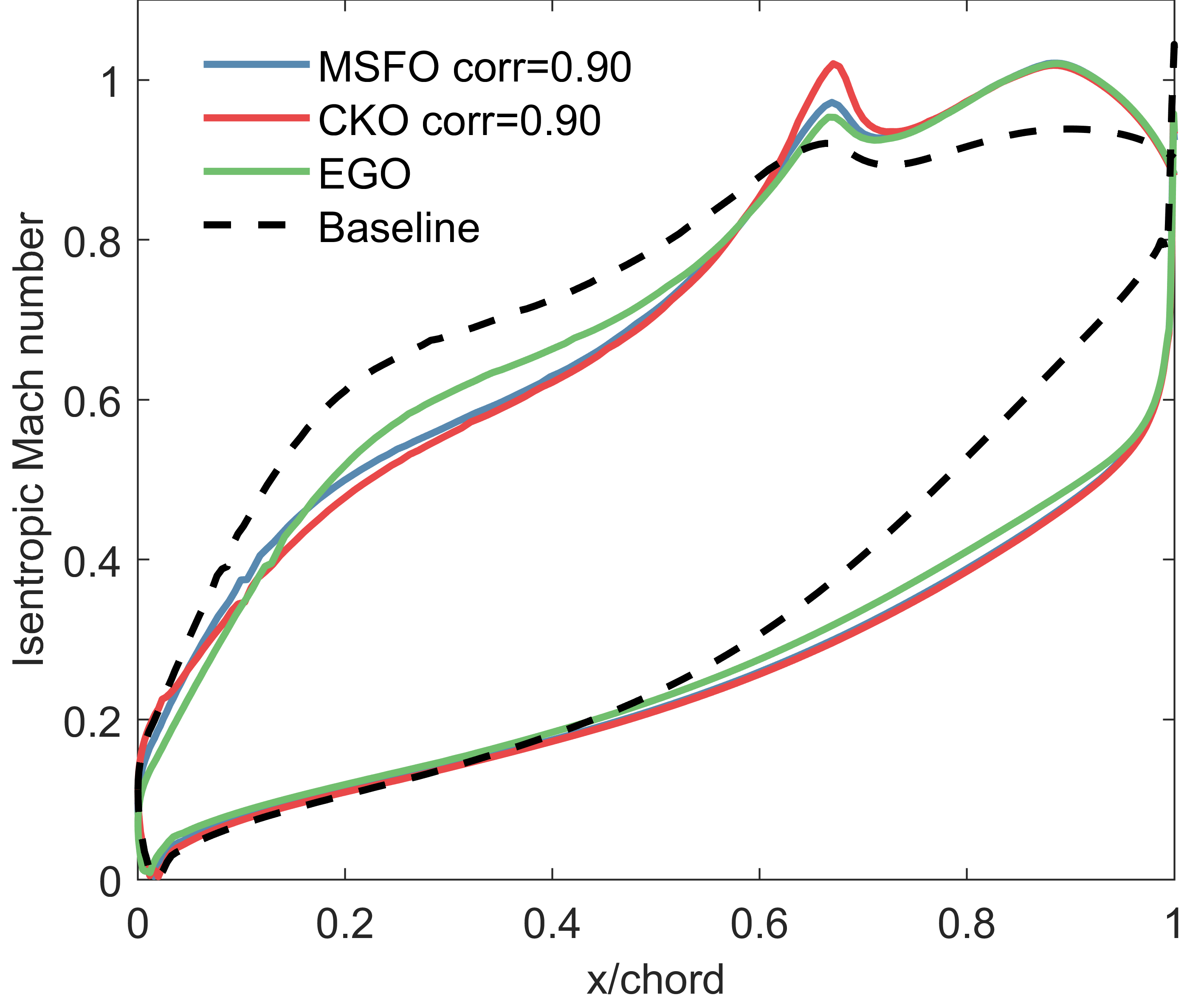}
\caption{Comparison of isentropic Mach number distribution on blade surface before and after GE-E3 optimization\label{fig:8}}
\end{figure}

After the MSFO algorithm is optimized with LF data of $corr=0.90$, the energy loss coefficient of blade profile is reduced from 3.98\% to 3.20\%, which is reduced by 0.78\%. 
Figure \ref{fig:8} shows the isentropic Mach number blade profile of the blade surface before and after optimization. It can be seen that the optimization results of different algorithms are similar. 

\begin{figure}[htbp]
\centering\includegraphics[width=0.9\linewidth]{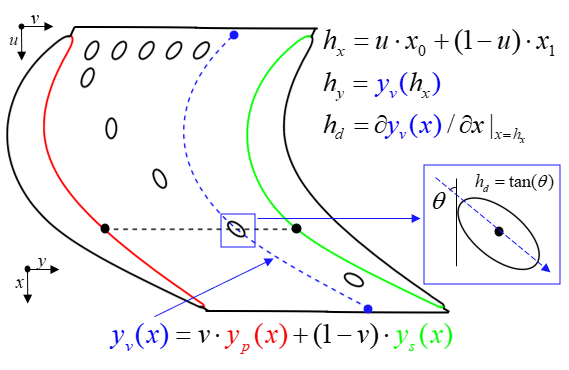}
\caption{The definition of parameters in the optimization of endwall cooling layout design\label{fig:9}}
\end{figure}
The optimized blade profiles are all post-loaded, which is beneficial to reduce aerodynamic loss and improve flow efficiency.
\subsection{Turbine endwall cooling layout design problem}
\par
The effectiveness of the MSFO algorithm was demonstrated through an expensive engineering problem, the turbine endwall cooling layout design problem.  
As shown in Figure \ref{fig:9}, $u$ and $v$ are the normalized axial and circumferential distances respectively. 
The cooling layout consists of a circumferential row and an axial row, each with five film cooling holes.
Since the relative coordinates $[u,v]$ of a row of holes are evenly distributed, the relative coordinates of five holes can be determined using only the coordinates of the first and last two holes.
Furthermore, the coordinates of some holes are fixed, so only 5 design variables, respectively ($v_{c1}$, $v_{c5}$, $v_{a1}$, $v_{a5}$ and $u_{a5}$) can controlling the layout of all film cooling holes. 
The ranges of above five design variables are shown in the Table \ref{tab:2}.
In Figure \ref{fig:9}, The virtual streamline $y_v(x)$ is obtained by weighted suction surface curve $y_s(x)$ and pressure surface curve $y_p(x)$. 
The position of the film cooling holes can be determined by the intersection of the line $x=C_{ax}\dot u$ and the virtual streamline $y_v(x)$. 
The angle of a film cooling hole is consistent with the tangent line of $y_v(x)$.
More information regarding design variables and boundary conditions can be found in Ref.\cite{buImprovingFilmCooling2022}.
\begin{table}[htbp]
    \centering
\caption{The definition of parameters in the optimization of endwall cooling layout design variables in the endwall cooling layout design case}\label{tab:2}
    \centering
    \begin{tabular}{ccc}
    \hline
    No. &Design Variable & Variation Range  \\ \hline
        1 &$v_{c1}$ &  [0.05,0.30]  \\ 
        2 &$v_{c5}$  & [0.70,0.93]  \\ 
        3 &$v_{a1}$  & [0.05,0.93]  \\ 
        4 &$v_{a5}$  & [0.05,0.93]  \\
        5 &$u_{a5}$ &  [0.55,0.90]  \\ \hline
    \end{tabular}
\end{table}
\par
\begin{figure}[htbp]
\begin{subfigure}[t]{0.25\textwidth} %
\centering{
\includegraphics[width=0.95\linewidth]{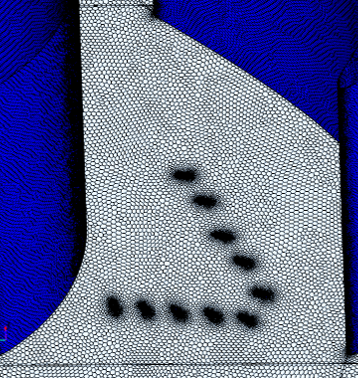}
}%
\subcaption{High-fidelity computation}
\end{subfigure}%
\begin{subfigure}[t]{0.25\textwidth}
\centering{%
\includegraphics[width=0.95\linewidth]{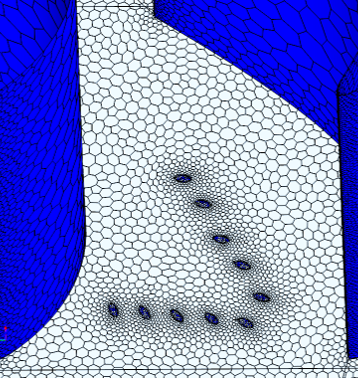}
}%
\subcaption{Low-fidelity computation}
\end{subfigure}
\caption{Meshes for high- and low-fidelity computations in the optimization of endwall cooling layout design. }\label{fig:10}
\end{figure}

In the optimization of turbine endwall cooling layout design problem, HF and LF data are also derived from CFD calculations with fine and coarse meshes, respectively. 
The two meshes are shown in Fig.\ref{fig:10}, where the HF mesh has about $6\times 10^{7}$ nodes, while the LF mesh has only $1/8$ nodes of the former. 
And the correlation coefficient between the CFD results with two kinds of meshes is 0.87 in this problem.
The specific setting and verification of CFD simulation can also be referred to literature~\cite{buImprovingFilmCooling2022}.
In this optimization design, the optimization target is to minimize the overheating area $A_h$, that is, the area of the region with the overall cooling effectiveness $\phi$ less than 0.15.
So, the definition of the optimization target in the turbine endwall cooling layout design problem is expressed in Eq.(\ref{eq:4.3}). 
\begin{equation}\label{eq:4.3}
    \begin{split}
    \text{min  } \{A_h({\bf{x}})\} \\
\phi=\frac{T_{\infty}-T_w}{T_{\infty}-T_{c, in}}\\
A_h=\iint_{s: \phi<0.15} ds\\
\end{split}
\end{equation}
\par
\begin{figure}[htbp]
\centering\includegraphics[width=0.7\linewidth]{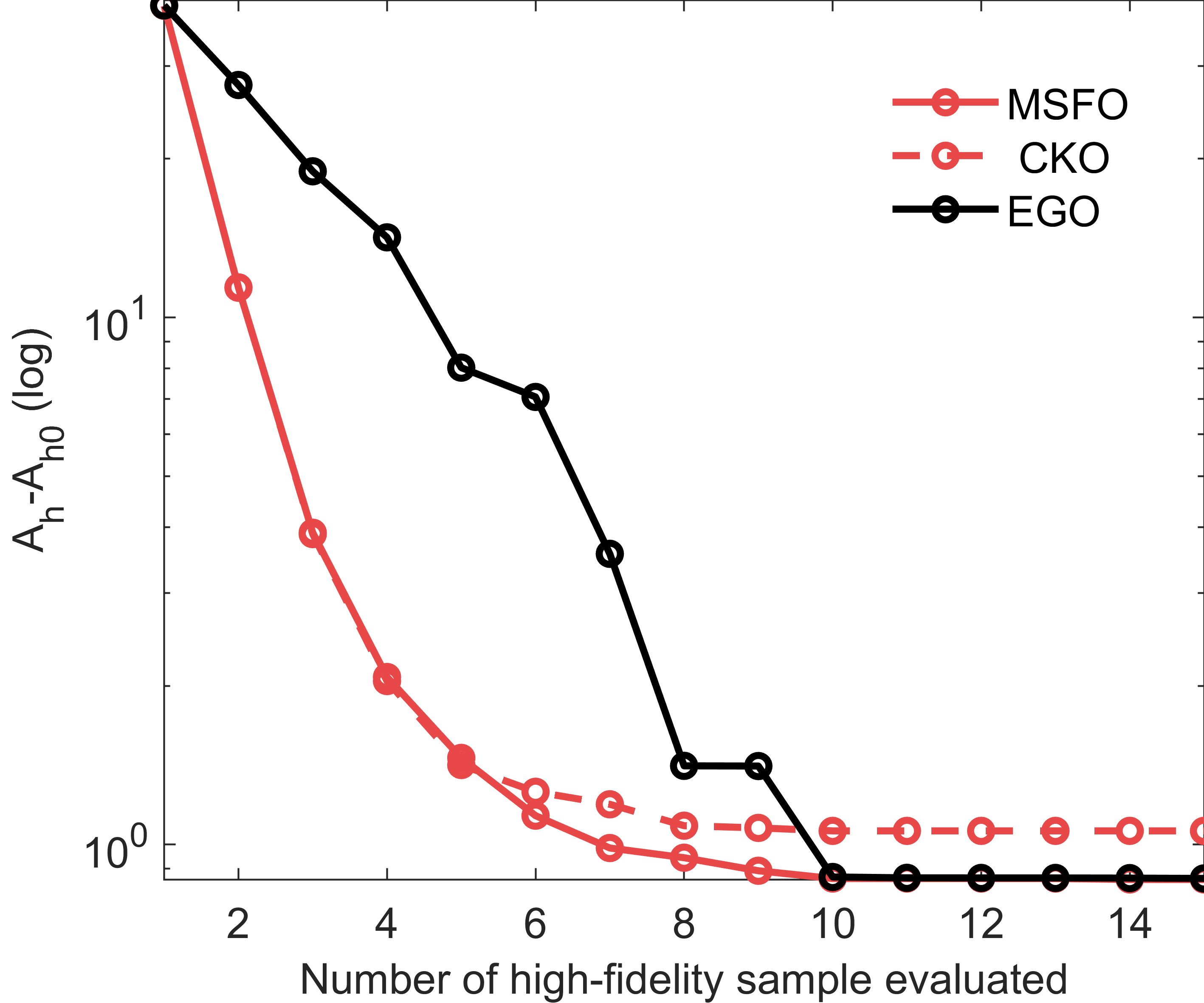}
\caption{The average convergence curves of the endwall optimization task\label{fig:11}}
\end{figure}

The optimization utilized the MSFO, CKO, and EGO algorithms. 
Figure \ref{fig:11} illustrates the convergence history of the three algorithms, with the overheating area target denoted as $A_{h0}=111.0\text{ cm}^2$. 
During the initial stages, both the MSFO and CKO algorithms converged rapidly, with their convergence curves being almost identical. However, by the sixth iteration, the convergence rate of CKO decreased substantially and stagnated, resulting in a worse optimal value than EGO. 
On the other hand, the MSFO algorithm maintained a faster convergence rate and outperformed the EGO algorithm during the entire optimization process.
\par
\begin{figure}[htbp]
    \begin{subfigure}[t]{0.5\textwidth} %
    \centering{
    \includegraphics[width=0.45\linewidth]{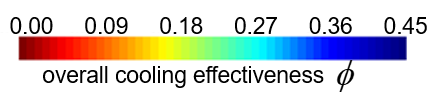}
    }%
    \subcaption*{ }
    \end{subfigure}%

    \begin{subfigure}[t]{0.5\textwidth}
    \centering{%
    \includegraphics[width=0.65\linewidth]{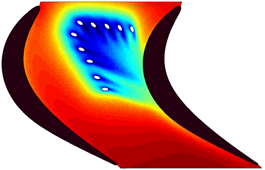}
    }%
    \subcaption{Baseline design}
    \end{subfigure}

    \begin{subfigure}[t]{0.5\textwidth}
    \centering{%
    \includegraphics[width=0.65\linewidth]{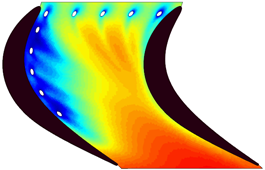}
    }%
    \subcaption{Optimal design of the MSFO}
    \end{subfigure}

    \caption{Comparison of end-wall cooling efficiency distribution before and after optimization}\label{fig:12}
    \end{figure}

 
Figure \ref{fig:12} shows the distribution of cooling effectiveness $\phi$ on end wall before and after optimization. 
Through optimization, the endwall overheating area $A_h$ was reduced from $324.0 \text{ cm}^2$ to $111.8 \text{ cm}^2$.
It can be seen that after optimization, the circumferential row cooling holes move to the front edge, and their distances between each other become larger. 
The axial row moves towards the pressure side and forms a large area of high cooling effectiveness beside the pressure surface of blade. 
It's obvious that the optimized distribution effectively increases the cooling effectiveness of the endwall surface.
\par
The practical applicability of the MSFO algorithm in real-world engineering has been demonstrated through the aforementioned two engineering optimizations. 
Test results indicate that the local search capability of the MSFO method has been significantly enhanced compared to the CKO algorithm. Furthermore, compared to the EGO algorithm, the MSFO algorithm exhibits significantly faster initial convergence and achieves superior results in later stages.

\section{Conclusion}
In this work, a simple analysis of the reasons for the insufficient local search ability in the late stage of MFO algorithm was made.
In the conventional MFS, the local adjustment of the model cannot be realized only through the scale factor $\rho$, so the global accuracy is improved while the local accuracy is decreased in the case of uneven distribution of samples.
\par
In this study, a comprehensive analysis was conducted to identify the underlying causes for the insufficient local search ability observed in the late stage of the multi-fidelity optimization (MFO) algorithm. 
It was observed that in the conventional MFS, the local adjustment of the model solely through the scale factor $\rho$ is insufficient, leading to improved global accuracy but reduced local accuracy, especially in scenarios with uneven sample distribution.
\par
To address this limitation of the MFO algorithm, an ensembled multi-fidelity surrogate (EMFS) was proposed, which enables local weight adjustments and operates as follows:
(1) Distinguishing local high-density high-fidelity samples using the DBSCAN method.
(2) Determining the weight coefficients of single- and multi-fidelity surrogates in the design space.
(3) Establishing the ensemble multi-fidelity surrogate.
Building upon this, the multi-single-fidelity optimization (MSFO) algorithm was formulated to tackle expensive black-box optimization problems with multiple sample sources. 
The MSFO method was demonstrated on numerical cases and two engineering problems related to GE-E3 blade aerodynamic design and turbine endwall cooling layout design, and was compared with the current MFO algorithm. 
The results indicate that the MSFO algorithm efficiently optimizes multi-fidelity problems, and successfully achieves two main objectives. 
On one hand, the MSFO algorithm exhibits robust local search ability, avoiding stagnation in the later stage of optimization. 
On the other hand, the MSFO algorithm demonstrates strong robustness in handling misleading low-fidelity information. 
The findings of this study highlight the significant potential of the MSFO algorithm in the field of turbomachinery design.

\section*{Acknowledgments}
The authors would like to thank the anonymous referees for their valuable comments. 
This work was supported by the National Science and Technology Major Project (2019-II-0008-0028), the Industry-University-Research Cooperation Project of Aero Engine Corporation of China (HFZL2021CXY004) and the High-level Innovative and Entrepreneurial Talents Introduction Project of Qinchuangyuan (QCYRCXM-2022-210).

\bibliographystyle{asmeconf}  %
\bibliography{bibfile}%

\begin{thebibliography}{10}
\newcommand{\enquote}[1]{``#1''}
\providecommand{\url}[1]{\texttt{#1}}
\providecommand{\urlprefix}{URL }
\expandafter\ifx\csname urlstyle\endcsname\relax
  \providecommand{\doi}[1]{DOI \discretionary{}{}{}#1}\else
  \providecommand{\doi}{DOI \discretionary{}{}{}\begingroup
  \urlstyle{rm}\Url}\fi
\providecommand{\eprint}[2][]{\urlprefix\url{#1#2}}

\bibitem{songResearchMetamodelBasedGlobal2016}
Song, Liming, Guo, Zhendong, Li, Jun and Feng, Zhenping.
\newblock \enquote{Research on {{Metamodel-Based Global Design Optimization}}
  and {{Data Mining Methods}}.}
\newblock \textit{Journal of Engineering for Gas Turbines and Power} Vol. 138
  No.~9 (2016): p. 092604.

\bibitem{ruanVariablefidelityProbabilityImprovement2020}
Ruan, Xiongfeng, Jiang, Ping, Zhou, Qi, Hu, Jiexiang and Shu, Leshi.
\newblock \enquote{Variable-Fidelity Probability of Improvement Method for
  Efficient Global Optimization of Expensive Black-Box Problems.}
\newblock \textit{Structural and Multidisciplinary Optimization}  (2020).

\bibitem{songOptimizationKnowledgeDiscovery2018}
Song, Liming, Guo, Zhendong, Li, Jun and Feng, Zhenping.
\newblock \enquote{Optimization and {{Knowledge Discovery}} of a
  {{Three-Dimensional Parameterized Vane}} with {{Nonaxisymmetric Endwall}}.}
\newblock \textit{Journal of Propulsion and Power} Vol.~34 No.~1 (2018): pp.
  234--246.

\bibitem{liuSurveyAdaptiveSampling2017}
Liu, Haitao, Ong, Yew-Soon and Cai, Jianfei.
\newblock \enquote{A Survey of Adaptive Sampling for Global Metamodeling in
  Support of Simulation-Based Complex Engineering Design.}
\newblock \textit{Structural and Multidisciplinary Optimization} Vol.~57 No.~1
  (2017): pp. 393--416.

\bibitem{liuSequentialSamplingGeneration2021a}
Liu, Yin, Li, Kunpeng, Wang, Shuo, Cui, Peng, Song, Xueguan and Sun, Wei.
\newblock \enquote{A {{Sequential Sampling Generation Method}} for
  {{Multi-Fidelity Model Based}} on {{Voronoi Region}} and {{Sample Density}}.}
\newblock \textit{Journal of Mechanical Design} Vol. 143 No.~12 (2021): p.
  121702.

\bibitem{jonesEfficientGlobalOptimization}
Jones, Donald~R and Schonlau, Matthias.
\newblock \enquote{Efficient {{Global Optimization}} of {{Expensive Black-Box
  Functions}}.}  : p.~38.

\bibitem{forresterMultifidelityOptimizationSurrogate2007}
Forrester, Alexander~I.J, S{\'o}bester, Andr{\'a}s and Keane, Andy~J.
\newblock \enquote{Multi-Fidelity Optimization via Surrogate Modelling.}
\newblock \textit{Proceedings of the Royal Society A: Mathematical, Physical
  and Engineering Sciences} Vol. 463 No. 2088 (2007): pp. 3251--3269.

\bibitem{makkarMachineLearningFramework2022}
Makkar, Gaurav, Smith, Cameron, Drakoulas, George, Kopsaftopoulos, Fotis and
  Gandhi, Farhan.
\newblock \enquote{A {{Machine Learning Framework}} for {{Physics-Based
  Multi-Fidelity Modeling}} and {{Health Monitoring}} for a {{Composite
  Wing}}.}
\newblock \textit{Volume 1: {{Acoustics}}, {{Vibration}}, and {{Phononics}}}:
  p. V001T01A008. 2022. {American Society of Mechanical Engineers}, {Columbus,
  Ohio, USA}.

\bibitem{parkRemarksMultifidelitySurrogates2017}
Park, Chanyoung, Haftka, Raphael~T. and Kim, Nam~H.
\newblock \enquote{Remarks on Multi-Fidelity Surrogates.}
\newblock \textit{Structural and Multidisciplinary Optimization} Vol.~55 No.~3
  (2017): pp. 1029--1050.

\bibitem{shiMultiFidelityModelingAdaptive2020}
Shi, Renhe, Liu, Li, Long, Teng, Wu, Yufei and Gary~Wang, G.
\newblock \enquote{Multi-{{Fidelity Modeling}} and {{Adaptive Co-Kriging-Based
  Optimization}} for {{All-Electric Geostationary Orbit Satellite Systems}}.}
\newblock \textit{Journal of Mechanical Design} Vol. 142 No.~2 (2020): p.
  021404.

\bibitem{benamaraMultifidelityPODSurrogateassisted2017}
Benamara, Tariq, Breitkopf, Piotr, Lepot, Ingrid, Sainvitu, Caroline and
  Villon, Pierre.
\newblock \enquote{Multi-Fidelity {{POD}} Surrogate-Assisted Optimization:
  {{Concept}} and Aero-Design Study.}
\newblock \textit{Structural and Multidisciplinary Optimization} Vol.~56 No.~6
  (2017): pp. 1387--1412.

\bibitem{mondalMultiFidelityGlobalLocalOptimization2019}
Mondal, Sudeepta, Joly, Michael~M. and Sarkar, Soumalya.
\newblock \enquote{Multi-{{Fidelity Global-Local Optimization}} of a
  {{Transonic Compressor Rotor}}.}
\newblock \textit{Volume {{2D}}: {{Turbomachinery}}}: p. V02DT46A020. 2019.
  {American Society of Mechanical Engineers}, {Phoenix, Arizona, USA}.

\bibitem{linSequentialSamplingApproach2022a}
Lin, Quan, Zhou, Qi, Hu, Jiexiang, Cheng, Yuansheng and Hu, Zhen.
\newblock \enquote{A {{Sequential Sampling Approach}} for {{Multi-Fidelity
  Surrogate Modeling-Based Robust Design Optimization}}.}
\newblock \textit{Journal of Mechanical Design} Vol. 144 No.~11 (2022): p.
  111703.

\bibitem{guoGenerativeMultiformBayesian2022}
Guo, Zhendong, Liu, Haitao, Ong, Yew-Soon, Qu, Xinghua, Zhang, Yuzhe and Zheng,
  Jianmin.
\newblock \enquote{Generative {{Multiform Bayesian Optimization}}.}
\newblock \textit{IEEE Transactions on Cybernetics}  (2022): pp. 1--14.

\bibitem{wangTransferOptimizationAccelerating2020a}
Wang, Qineng, Song, Liming, Guo, Zhendong and Li, Jun.
\newblock \enquote{Transfer {{Optimization}} in {{Accelerating}} the {{Design}}
  of {{Turbomachinery Cascades}}.}
\newblock \textit{Proceedings of {{ASME Turbo Expo}} 2020 {{Turbomachinery
  Technical Conference}} and {{Exposition}}}: p.~12. 2020.

\bibitem{gisellefernandez-godinoIssuesDecidingWhether2019}
{Giselle Fern{\'a}ndez-Godino}, M., Park, Chanyoung, Kim, Nam~H. and Haftka,
  Raphael~T.
\newblock \enquote{Issues in {{Deciding Whether}} to {{Use Multifidelity
  Surrogates}}.}
\newblock \textit{AIAA Journal} Vol.~57 No.~5 (2019): pp. 2039--2054.

\bibitem{guoAnalysisDatasetSelection2018}
Guo, Zhendong, Song, Liming, Park, Chanyoung, Li, Jun and Haftka, Raphael~T.
\newblock \enquote{Analysis of Dataset Selection for Multi-Fidelity Surrogates
  for a Turbine Problem.}
\newblock \textit{Structural and Multidisciplinary Optimization} Vol.~57 No.~6
  (2018): pp. 2127--2142.

\bibitem{zhouGeneralizedHierarchicalCoKriging2020}
Zhou, Qi, Wu, Yuda, Guo, Zhendong, Hu, Jiexiang and Jin, Peng.
\newblock \enquote{A Generalized Hierarchical Co-{{Kriging}} Model for
  Multi-Fidelity Data Fusion.}
\newblock \textit{Structural and Multidisciplinary Optimization}  (2020).

\bibitem{parkLowfidelityScaleFactor2018}
Park, Chanyoung, Haftka, Raphael~T. and Kim, Nam~H.
\newblock \enquote{Low-Fidelity Scale Factor Improves {{Bayesian}}
  Multi-Fidelity Prediction by Reducing Bumpiness of Discrepancy Function.}
\newblock \textit{Structural and Multidisciplinary Optimization} Vol.~58 No.~2
  (2018): pp. 399--414.

\bibitem{shuNovelApproachSelecting2019}
Shu, Leshi, Jiang, Ping, Song, Xueguan and Zhou, Qi.
\newblock \enquote{Novel {{Approach}} for {{Selecting Low-Fidelity Scale
  Factor}} in {{Multifidelity Metamodeling}}.}
\newblock \textit{AIAA Journal} Vol.~57 No.~12 (2019): pp. 5320--5330.

\bibitem{buSelectingScaleFactor2022}
Bu, Hongyan, Song, Liming, Guo, Zhendong and Li, Jun.
\newblock \enquote{Selecting Scale Factor of {{Bayesian}} Multi-Fidelity
  Surrogate by Minimizing Posterior Variance.}
\newblock \textit{Chinese Journal of Aeronautics}  (2022): p.
  S1000936122001042.

\bibitem{schubertDBSCANRevisitedRevisited2017}
Schubert, Erich, Sander, J{\"o}rg, Ester, Martin, Kriegel, Hans~Peter and Xu,
  Xiaowei.
\newblock \enquote{{{DBSCAN Revisited}}, {{Revisited}}: {{Why}} and {{How You
  Should}} ({{Still}}) {{Use DBSCAN}}.}
\newblock \textit{ACM Transactions on Database Systems} Vol.~42 No.~3 (2017):
  pp. 1--21.

\bibitem{forresterRecentAdvancesSurrogatebased2009}
Forrester, Alexander~I.J. and Keane, Andy~J.
\newblock \enquote{Recent Advances in Surrogate-Based Optimization.}
\newblock \textit{Progress in Aerospace Sciences} Vol.~45 No. 1-3 (2009): pp.
  50--79.

\bibitem{jonesTaxonomyGlobalOptimization}
Jones, Donald~R.
\newblock \enquote{A {{Taxonomy}} of {{Global Optimization Methods Based}} on
  {{Response Surfaces}}.}  : p.~39.

\bibitem{kennedyPredictiwnghetnheFaOsuttApuptprforoxmimaatCioonmsParleexACvoamilapbulteerCode}
Kennedy, M~C and O'Hagan, A.
\newblock \enquote{{{PredictiwnghetnheFaOsuttApuptprforoxmimaatCioonms
  parleexACvoamilapbulteer Code}}.}  : p.~17.

\bibitem{zerpaOptimizationMethodologyAlkaline2005}
Zerpa, Luis~E., Queipo, Nestor~V., Pintos, Salvador and Salager, Jean-Louis.
\newblock \enquote{An Optimization Methodology of Alkaline\textendash
  Surfactant\textendash Polymer Flooding Processes Using Field Scale Numerical
  Simulation and Multiple Surrogates.}
\newblock \textit{Journal of Petroleum Science and Engineering} Vol.~47 No. 3-4
  (2005): pp. 197--208.

\bibitem{huangSequentialKrigingOptimization2006}
Huang, D., Allen, T.~T., Notz, W.~I. and Miller, R.~A.
\newblock \enquote{Sequential Kriging Optimization Using Multiple-Fidelity
  Evaluations.}
\newblock \textit{Structural and Multidisciplinary Optimization} Vol.~32 No.~5
  (2006): pp. 369--382.

\bibitem{bertholdClusteringTimeSeries2016}
Berthold, Michael~R. and H{\"o}ppner, Frank.
\newblock \enquote{On {{Clustering Time Series Using Euclidean Distance}} and
  {{Pearson Correlation}}.} (2016).

\bibitem{guoParallelMultifidelityExpected2021}
Guo, Zhendong, Wang, Qineng, Song, Liming and Li, Jun.
\newblock \enquote{Parallel Multi-Fidelity Expected Improvement Method for
  Efficient Global Optimization.}
\newblock \textit{Structural and Multidisciplinary Optimization} Vol.~64 No.~3
  (2021): pp. 1457--1468.

\bibitem{buImprovingFilmCooling2022}
Bu, Hongyan, Yang, Yufeng, Song, Liming and Li, Jun.
\newblock \enquote{Improving the {{Film Cooling Performance}} of a {{Turbine
  Endwall With Multi-Fidelity Modeling Considering Conjugate Heat Transfer}}.}
\newblock \textit{Journal of Turbomachinery} Vol. 144 No.~1 (2022): p. 011011.

\end{thebibliography}

\appendix

\end{document}